\documentclass{article}
\usepackage{spconf,amsmath,graphicx}
\usepackage{multirow}
\usepackage{setspace}
\usepackage{algorithm}
\usepackage{algorithmic}
\usepackage{epstopdf}
\usepackage{balance}

\title{Cascade RNN-Transducer: Syllable Based Streaming On-device Mandarin Speech Recognition with a Syllable-to-Character Converter}
\name{Xiong Wang$^{*}$\thanks{The first three authors contributed equally to this work. This paper is supported by MoE-CMCC “Artifical Intelligence” Project (MCM20190701).}, Zhuoyuan Yao$^{*}$, Xian Shi$^{*}$, Lei Xie\sthanks{Lei Xie is the corresponding author.}}
\address{Audio, Speech and Language Processing Group (ASLP@NPU),\\
	School of Computer Science, Northwestern Polytechnical University, Xi'an, China}
\begin{document}
%
\maketitle
\begin{abstract}
End-to-end models are favored in automatic speech recognition (ASR) because of its simplified system structure and superior performance. Among these models, recurrent neural network transducer (RNN-T) has achieved significant progress in streaming on-device speech recognition because of its high-accuracy and low-latency. RNN-T adopts a prediction network to enhance language information, but its language modeling ability is limited because it still needs paired speech-text data to train. Further strengthening the language modeling ability through extra text data, such as shallow fusion with an external language model, only brings a small performance gain. In view of the fact that Mandarin Chinese is a character-based language and each character is pronounced as a tonal syllable, this paper proposes a novel cascade RNN-T approach to improve the language modeling ability of RNN-T. Our approach firstly uses an RNN-T to transform acoustic feature into syllable sequence, and then converts the syllable sequence into character sequence through an RNN-T-based syllable-to-character converter. Thus a rich text repository can be easily used to strengthen the language model ability. By introducing several important tricks, the cascade RNN-T approach surpasses the character-based RNN-T by a large margin on several Mandarin test sets, with much higher recognition quality and similar latency.
\end{abstract}
\section{Introduction}\label{sec1}

Conventional automatic speech recognition (ASR) usually adopts a hybrid system of deep neural network - hidden Markov model (DNN-HMM)~\cite{hinton2012deep} which is complex and requires a considerable amount of computing resource, so it is difficult to deploy on edge devices. Recently, end-to-end (E2E) speech recognition has achieved significant progress with simplified system architecture and superior performance. The E2E models usually adopt a sequence-to-sequence (S2S) framework to directly transform acoustic feature sequences into text sequences through specifically-designed neural networks. These models are particularly favored on edge devices for more concise architecture and reduced computing resource consumption over the hybrid ASR systems. However, the E2E speech recognition models, modeling acoustic and language information jointly in a unified framework, usually require a large amount of paired speech-text data for model training. Thus it is difficult for the models themselves to acquire strong language modeling ability by using available text data with more orders of magnitude than the speech-text paired data, especially when the training set does not match the language domain of specific applications. This paper addresses this problem by introducing a novel approach to improve the language modeling ability of E2E models.

As an S2S model, recurrent neural network transducer (RNN-T)~\cite{graves2012sequence} and its variants have achieved high-accuracy and low-latency in streaming on-device speech recognition~\cite{rnntshibie,graves2013speech}. Neural transducer has the streaming decoding ability in nature, while other E2E competitors, particularly those based on attention mechanism, such as transformer~\cite{transformer,povey2018time,dong2018speech} and listen, attend and spell~\cite{las} (LAS), have to be modified to possess the streaming ability. RNN-T uses a prediction network to enhance the language information based on the connectionist temporal classification (CTC)~\cite{graves2014towards} criterion. But its language modeling ability is not satisfactory because it still needs paired speech-text data to train. A recent study has unveiled that the language modeling ability of the prediction network is still quite weak~\cite{ghodsi2020rnn}.

Plenty of effort has been made on improving the performance of E2E models by introducing additional language information. A common solution is to use a language model (LM) fusion strategy: an LM is first externally trained on text data and then incorporated into the E2E model~\cite{shallowfusion,zhao2019shallow}. Shallow fusion simply interpolates the label probabilities with the ones from an external LM during decoding stage. Other fusion variants, such as deep fusion, cold fusion and component fusion, have also been proposed. Data augmentation through speech synthesis is another solution. Work in ~\cite{ttsrnnt,ttsrnnt2} has shown that data augmentation with text-to-speech utterances yields improvement to E2E models; however, there still remains a substantial gap in performance between models trained on human speech and those trained on synthesized speech. Similar to the tricks used in conventional hybrid approaches, two-pass decoding also can be introduced to E2E models with improved recognition performance. Recently, a two-pass RNN-T+LAS model, where LAS rescores hypotheses from RNN-T, has been proposed~\cite{rnntlas} and improved further with more tricks~\cite{rnntlas2}. To surpass server-side conventional model, trade-off between quality and latency has been particularly considered.

Most approaches on neural transducer have been conducted on English corpora and different modeling units, such as phonemes, grapheme and word-piece, have been explored~\cite{rao2017exploring}. In this paper, we are particularly interested in streaming on-device Mandarin ASR using RNN-T. Mandarin Chinese is significantly different from English in both written and spoken aspects. Chinese is a character-based language and each character is pronounced as a tonal syllable. There are several studies on LAS and Transformer based Mandarin ASR, but we only find one paper on the use of RNN-T in Mandarin which shows its feasibility on modeling Chinese characters~\cite{sengmaornnt}. Character has been widely chosen as a natural modeling unit in Transformer-based Mandarin ASR as well. However, as Chinese has a huge set of characters, all these works have chosen a partial set of frequently-used characters to model while the rest are simply abandoned, which means the abandoned characters can never be outputted, leading to out-of-vocabulary (OOV) problem.

In this paper, we propose a novel \textit{cascade RNN-T} approach for streaming on-device Mandarin speech recognition. Specifically, we cascade two RNN transducers to strengthen the language modeling ability -- the first transforms acoustic input into a syllable sequence, while the second converts the syllable sequence into the final character sequence. The proposed approach has the following advantages: 1) a rich text repository can be easily used to strengthen the language modeling ability; 2) the OOV issue does not exist by the introduction of RNN-T based syllable-to-character (S2C) converter; 3) streaming ability has been maintained as the use of the transducer framework. By introducing several important tricks on the proposed syllable-based cascade RNN-T, including adding convolution layer, self shallow fusion, text augmentation and syllable correction, we manage to surpass the character-based RNN-T by a large margin. Compared with character RNN-T with shallow fusion, cascade RNN-T has an obvious improvement on several Mandarin test sets, with higher recognition quality and similar latency.

\section{RNN-Transducer}\label{sec2}
Our proposed approach is based on a modified version of oracle RNN-T~\cite{graves2012sequence}. Here we first introduce the RNN-T structure as well as the typical shallow fusion strategy to strengthen the language modeling ability.

\begin{figure}
	\centering
	\includegraphics[width=1\linewidth]{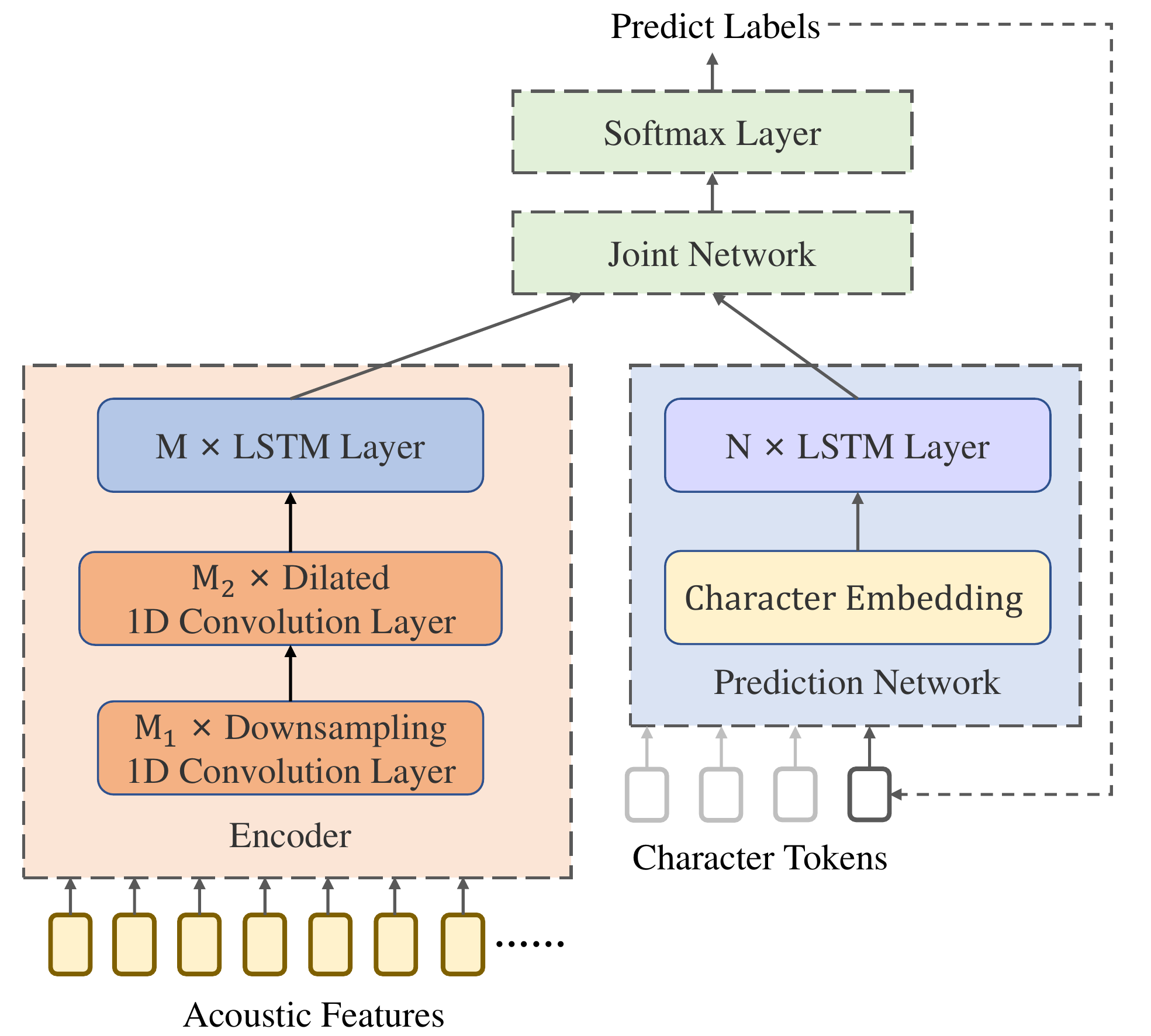}\vspace{0pt}
	\caption{The RNN-T architecture used in this paper.}\vspace{-0pt}
	\label{fig:1}
\end{figure}

\subsection{Model architecture}
RNN-T has the ability to model the alignment between speech features $\mathbf{x}=[{{\mathbf{x}}_{1}},{{\mathbf{x}}_{2}},\cdots ,{{\mathbf{x}}_{T}}]$ and output label sequences $\mathbf{y}=[{{y}_{1}},{{y}_{2}},\cdots ,{{y}_{U}}]$, where $T$ is the number of feature frames and $U$ is the length of the output label sequence. 
Fig.~\ref{fig:1} shows the RNN-T structure used in this paper, which mainly includes an encoder, a predict network and a joint network. 
Specifically, the encoder converts the speech features into a high-dimensional representation ${\mathbf{h}}^{enc}$. Different from the oracle RNN-T~\cite{graves2012sequence}, we add $M_1$ layers of 1-D convolution with a stride size greater than 1 for down-sampling. This operation will effectively reduce the resource consumption during the training process and decoding stage. In addition, the $M_2$ layers of dilated 1-D convolution followed can capture local context information more effectively at a small cost of a short latency. The whole process can be expressed as
\begin{equation}
{{\mathbf{h}}^{enc}}=\text{Encoder}(\mathbf{x}).
\end{equation}
The main purpose of prediction network is to generate a higher-dimensional representation $h_{u}^{pred}$ of last label ${y}_{u-1}$ as shown in Eq.~(\ref{eq2}). In order to avoid the sparsity caused by directly using one-hot label as input, an embedding layer is arranged before $N$ layers of LSTM.
\begin{equation}\label{eq2}
h_{u}^{pred}=\text{Prediction}({{y}_{u-1}})
\end{equation}
The joint network is represented by several fully connected layers, and finally a softmax layer is used to predict the probability $P(k|t,u)$ of the next label, as shown in
\begin{equation}
h_{t,u}={{\mathbf{W}}^{joint}}\tanh (\mathbf{U}h_{t}^{enc}+\mathbf{V}h_{u}^{pred}+b)+{{b}_{joint}}
\end{equation}
\begin{equation}
P(k|t,u)=\text{softmax}(h_{t,u})
\end{equation}
where ${\mathbf{U}}$ and ${\mathbf{V}}$ is the projection matrix to combine $h_{t}^{enc}$ and $h_{u}^{pred}$, and ${\mathbf{W}}^{joint}$ is used to project network's output to the number of labels. During training, RNN-T uses the forward-backward algorithm~\cite{graves2013speech} to maximize the posterior of ${\bf{y}}$ given $\bf{x}$.

\subsection{Shallow fusion}
Shallow fusion~\cite{zhao2019shallow} adopts an RNNLM that can be trained using additional text data to improve the language modeling ability. The method is done by adding up the non-blank posterior probability of RNN-T and RNNLM in the logarithmic domain during the decoding process. As shown in Eq.~(\ref{eq5}), $\hat P({\bf{y}}|{\bf{x}})$ is the actual posterior probability used in the decoding process, $\log P({\bf{y}}|{\bf{x}})$ is the posterior probability of ${\bf{y}}$ given $\bf{x}$, and $\log P({\bf{y}})$ is the probability of ${\bf{y}}$ generated by RNNLM.
\begin{equation}\label{eq5}
\hat P({\bf{y}}|{\bf{x}}) = \log P({\bf{y}}|{\bf{x}}) + \lambda \log P({\bf{y}})
\end{equation}
When using shallow fusion, beam search can be adopted to ensure that as many decoding paths as possible are considered.

\section{Cascade RNN-Transducer}\label{sec3}
\subsection{Cascade model architecture}
\begin{figure}[!h]
	\centering
	\includegraphics[width=0.6\linewidth]{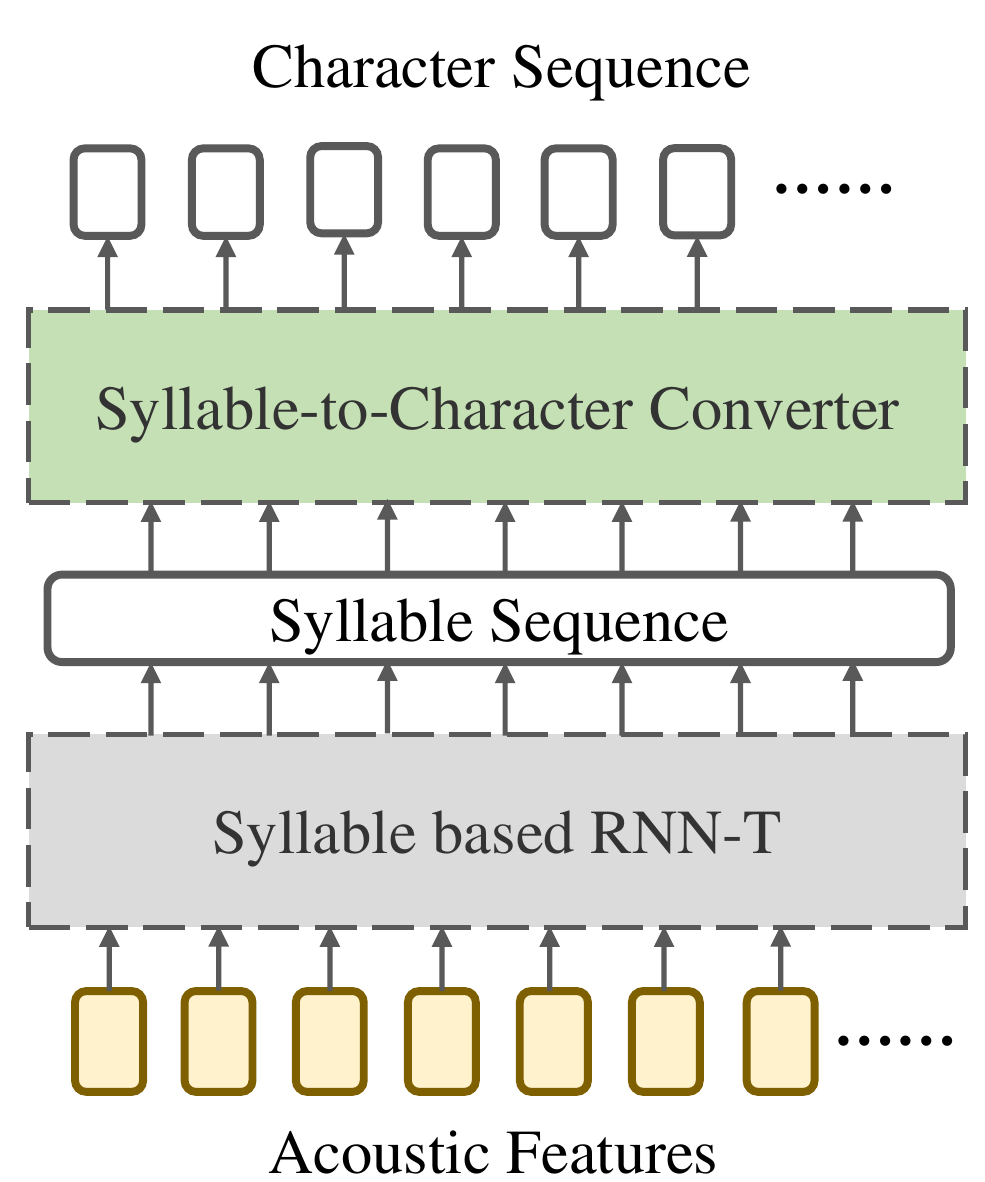}\vspace{0pt}
	\caption{The architecture of cascade RNN-T.}
	\label{fig:2}
\end{figure}

We propose a cascade RNN-T structure shown in Fig.~\ref{fig:2}. Formally in Eq.~(\ref{eq11}), the speech features $\mathbf{x}=[{{\mathbf{x}}_{1}},{{\mathbf{x}}_{2}},\cdots ,{{\mathbf{x}}_{T}}]$ first go through a syllable-level RNN-T, to obtain the syllable sequence $\mathbf{y}^{s}=[{{y}_{1}^{s},{{y}_{2}^{s}},\cdots ,{{y}_{U^s}^{s}}}]$, where $U^{s}$ is the length of the output syllable sequence.
\begin{equation}\label{eq11}
\mathbf{y}^{s} = \text{RNNT}_{Char}(\mathbf{x}).
\end{equation}
Then we use an S2C converter to convert the syllable sequence ${\mathbf{y}^{s}}$ into the character sequence $\mathbf{y}$:
\begin{equation}\label{eq12}
\mathbf{y} = \text{S2C}(\mathbf{y}^{s})
\end{equation}
During the training process, only the syllable-level RNN-T model in the first step needs data paired with speech and text, while the S2C converter in the second step can be trained only with text data. This is realized by another RNN-T.

\subsection{RNN-T based S2C Converter}
\begin{figure}[!h]
	\centering
	\includegraphics[width=1\linewidth]{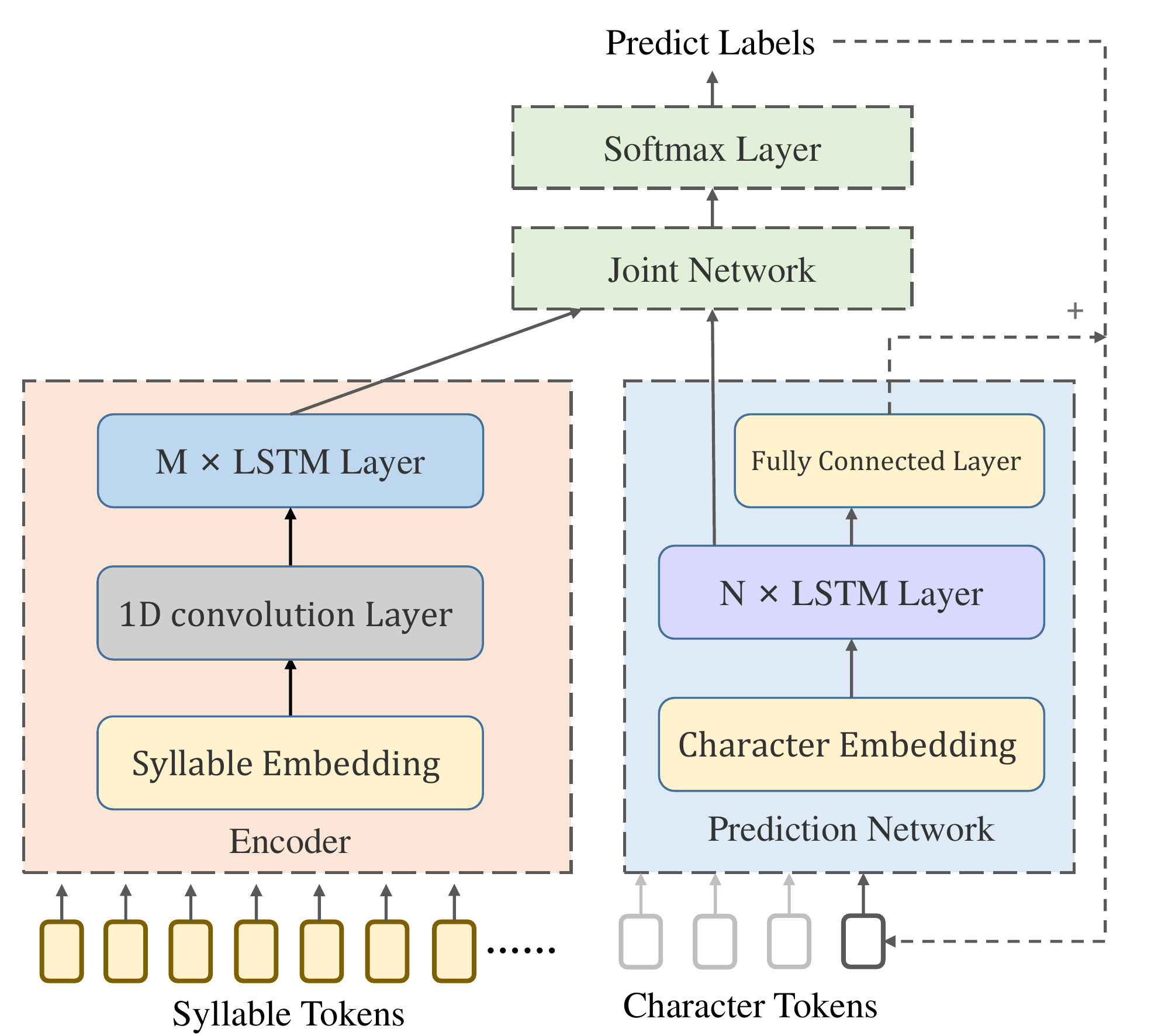}\vspace{0pt}
	\caption{RNN-T based syllable to character converter.}
	\label{fig:4}
\end{figure}

As shown in Fig.~\ref{fig:4}, the encoder input for the RNN-T based S2C converter is syllable sequence ${\mathbf{y}^{s}}$ while the input for the prediction network is character sequence $\mathbf{y}$. The output of joint network is the posterior distribution probability of the next character. Here we use an embedding layer to map one-hot input to high-dimensional representations in both encoder and prediction networks, and RNN-T loss is used in the training process as well.

\subsubsection{Convolution layer}
For a single LSTM layer, the network can only intuitively obtain the information of current token in each time step. In order to capture more context information, we add a convolution layer before the encoder, as described in
\begin{equation}
{{\mathbf{h}}^{conv}}=\text{Convolution}(\mathbf{y}^{s})
\end{equation} 
and
\begin{equation}
{{\mathbf{h}}^{enc}}=\text{Encoder}({{\mathbf{h}}^{conv}}).
\end{equation}
During training, in order to ensure the equal length of the input and the output of encoder, assuming that the kernel size of the convolution layer is $M_s$, $M_s-1$ zeros are padded at the start the syllable sequence.

\subsubsection{Self shallow fusion}
When using RNN-T as an S2C converter, a large amount of text data can be adopted to train the model. Hence we add a fully-connected layer after the prediction network to give the prediction network an additional task functioning as RNNLM and perform shallow fusion
\begin{equation}
\hat P({\bf{y}}|{\mathbf{y}^{s}}) = \log P({\bf{y}}|{\mathbf{y}^{s}}) + \lambda \log \text{FC}(\mathbf{h}^{pred}).
\end{equation}
We call this trick as \textit{self shallow fusion}. During the training process, we define the loss function as
\begin{equation}
\hat{L} = L_{\text{RNN-T}} + L_{ce}
\end{equation}
where $L_{RNN-T}$ is the loss of RNN-T and $L_{ce}$ is the loss of additional RNNLM task. The language modeling ability is implicitly embedded in the S2C RNN-T. But it still makes sense to use an additional task to explicitly do language modeling. The above multi-task training method can apparently reduce model parameters as well as computation as we do not need another RNNLM for shallow fusion -- this is why we call this trick as self shallow fusion.

\subsubsection{Text augmentation}
In order to prevent over-fitting, we adopt a text augmentation strategy, which is similar to SpecAugment~\cite{specaug} that is done to audio spectrum. Specifically, during training of S2C RNN-T, we randomly change several syllables in the syllable sequence to some other syllables as the input of the encoder. Alleviating over-fitting, this method can also be regarded as simulation on syllable errors caused by the first syllable-level RNN-T.

\subsubsection{Syllable correction}
The output of syllable-level RNN-T has insertion, deletion and replacement errors. This means that the input of S2C RNN-T inevitably has errors. So the S2C RNN-T desires the ability to convert the erroneous syllable sequence into character sequence that is grammatically and semantically correct. Inspired by recently studies~\cite{guo2019spelling,zhang2019automatic}, we use a syllable correction strategy to map the `noisy' syllable sequence to correct character sequence. Because RNN-T has the ability to map unequal length sequences in nature, we first decode the training set using the syllable-level RNN-T to syllable sequences. Then we use these syllable sequences with their corresponding correct character sequences to build additional correction text and mix these data with normal text data. At last, we use the mixed data to finetune an existing S2C RNN-T to improve the performance of the model.
\section{Experiments}\label{sec4}
\subsection{Dataset}
In this paper, we evaluated the proposed cascade RNN-T approach on two Mandarin speech recognition tasks: public AISHELL-2 corpus~\cite{aishell2} and internal 7,500 hours corpus. The AISHELL-2 corpus contains 1,000 hours of clean reading speech data collected from 1991 speakers~\footnote{Can be acquired from: www.aishelltech.com/aishell\_2}. The 7500-hour corpus contains reading speech data in fields such as entertainment, journalism, literature, technology and free conversation and so on. For both datasets, we reserve 5,000 sentences as development set. In addition, we use 2GB Chinese text data crawled from internet to train the RNNLM and the S2C converter. Character error rate (CER) is reported for AISHELL-1 test set (TA1)~\cite{aishell1}, AISHELL-2 test set (TA2)~\cite{aishell2}, an internal voice input test set (VI) and a voice assistant (VA) test set. The VI test set consists of about 3.4 hours data with 3,063 sentences covering many proper nouns and named entities, which is used to verify the language generalization ability of the speech recognition model. The VA test set consists of about 3.9 hours data with 5,000 speech commands to a voice assistant, which is challenging not only in language aspect, but also acoustically because speech is collected at various conditions some with low SNR.

\subsection{Experimental Setup}
For all experiments, the speech features are 71-dimensional log Mel-filterbank (FBank) computed on 25ms window with 10ms shift. We also applied SpecAugment~\cite{specaug} for acoustic data augmentation. For modeling units, we choose 5,139 characters for character-based models and 1,733 tonal syllables for syllable-based models. All experiments are conducted using TensorFlow~\cite{tensorflow2015-whitepaper} and Horovod~\cite{horovod}. During training process, we use random state passing (RSP) to avoid long-form problem~\cite{longform}. We adopt AdamOptimizer~\cite{adam} with learning rate at 0.0003 and gradient clipping at 5.0 for all models. Moreover, we employ layer normalization and variational recurrent dropout to prevent over-fitting. We use breadth-first beam search algorithm which is effective in exploring and combining alternative alignments~\cite{beamsearch}.

For the character RNN-T, the encoder network consists of 6 convolution layers and 5 LSTM layers while all convolution layers' kernel size is 3 and stride for each layer is \{2, 2, 1, 1, 1\}, while the number of filters is set to \{256, 256, 512, 512, 512\} and dilation rate is \{1, 1, 1, 2, 4\}. The prediction network has 2 LSTM layers and the joint network has 640 hidden units. For the cascade RNN-T, the first syllable-level RNN-T has the same architecture as the character RNN-T just mentioned. And for the second RNN-T for S2C conversion, the encoder network consists of 2 LSTM layers, the prediction network has a single LSTM layer and the joint network has 640 hidden units. The additional convolution layer has 1,024 filters and the kernel size is 3. All the LSTM layers mentioned above have 1,280 hidden units followed by a 640-dimensional projection layer. The RNNLM for shallow fusion in the character RNN-T approach consists of 2 LSTM layers each has 2,048 hidden units with a 640-dimensional projection layer.

\subsection{AISHELL-2 Task}
To verify our proposed cascade RNN-T, we first evaluate on the open AISHELL-2 corpus. As shown in Tab.~\ref{tab:1}, we obtain a small performance improvement from using beam search (B1) and shallow fusion (B2) for character-level RNN-T compared with B0 when the hyper-parameter $\lambda$ for shallow fusion is set to 0.35 and beam size is 5. For cascade RNN-T, during decoding process, we use greedy search for syllable-level RNN-T and beam search with beam size is 5 for S2C RNN-T. The performance of the cascade RNN-T (E0) is worse than character-level RNN-T. But we notice that the syllable error rate on TA2 for the syllable-level RNN-T in the cascade approach (E0) is 12.8\%, which is significantly lower than the CER of character RNN-T (B0) on the same test set. This phenomenon unveils that the syllable accuracy of the syllable-level RNN-T is higher than that of character-level RNN-T, so we believe that the second S2C RNN-T in the proposed cascade architecture holds great potential to be further strengthened to bring much better character error rate.

\begin{table}[!htb]
	\vspace{-10pt}
	\caption{Comparison of character-level RNN-T and Cascade RNN-T on test sets in CER.}
	\vspace{10pt}
	\label{tab:1}
	\centering
	\scalebox{0.8}{
		\begin{tabular}{|c|l|c|c|c|}
			\hline
			\rule[-1ex]{0pt}{3.5ex} \multirow{2}{*}{\textbf{Exp ID}} & \multirow{2}{*}{\textbf{Model}} & \multicolumn{3}{c|}{\textbf{CER (\%)}} \\
			\cline{3-5}	\rule[-1ex]{0pt}{3.5ex} && \textbf{TA1} & \textbf{TA2} & \textbf{VI} \\ \hline \hline
			\rule[-1ex]{0pt}{3.5ex} B0 & RNN-T Character & 6.71 & 16.27 & 21.02 \\ \hline
			\rule[-1ex]{0pt}{3.5ex} B1 & + Beam search & 6.55 & 15.79  & 20.44 \\ \hline
			\rule[-1ex]{0pt}{3.5ex} B2 & \ + Shallow fusion & \textbf{6.11} & \textbf{15.50} & \textbf{20.15} \\ \hline \hline
			\rule[-1ex]{0pt}{3.5ex} E0 & Cascade RNN-T & 17.4 & 22.08 & 24.52 \\ \hline
	\end{tabular}}
\end{table}

We further improve the S2C RNN-T followed the tricks introduced in Section 3.2, and the results are shown in Tab.~\ref{tab:2}. From the comparison between E1 and E0, it can be concluded that when the S2C RNN-T captures more context information through the convolution layer, the performance of cascade RNN-T can be significantly improved. For example, the CER on TA1 has been dramatically decreased from 17.4\% to 6.66\%. From E2 and E3, suppressing the overfitting of the model using text augmentation and explicitly introducing language information using self shallow fusion can also play a positive role. We use the decode result of syllable-level RNN-T on training set to generate correction text, and then finetune the model in E3 with the learning rate of 1e-4 while the normal text data is 7 times larger than correction text data in each batch. And finally, syllable correction (E4) can bring further performance gain and we manage to surpass the character-based RNN-T (B2) on all testing sets. Specifically on the challenging VI test set, we can achieve 12.65\% relative CER reduction. Tab.~\ref{tab:3} gives a detailed illustration the changes of deletion, insertion and substitute errors for the proposed tricks on TA2. From E0 to E4, we can see that all kinds of tricks we use can effectively reduce the substitute errors. At the same time, it also confirms that our proposed cascade RNN-T can greatly improve the language modeling ability of the model because substitute errors often represent grammatical errors.
\begin{table}[!htb]
	\vspace{-10pt}
	\caption{Comparison of the results of various tricks to improve the performance of cascade RNN-T on test sets in CER.}
	\vspace{10pt}
	\label{tab:2}
	\centering
	\scalebox{0.8}{
		\begin{tabular}{|c|l|c|c|c|}
			\hline
			\rule[-1ex]{0pt}{3.5ex} \multirow{2}{*}{\textbf{Exp ID}} & \multirow{2}{*}{\textbf{Model}} & \multicolumn{3}{c|}{\textbf{CER (\%)}} \\
			\cline{3-5} \rule[-1ex]{0pt}{3.5ex} && \textbf{TA1} & \textbf{TA2} & \textbf{VI} \\ \hline \hline
			\rule[0ex]{0pt}{0ex}   & RNN-T Character &   &   &   \\
			\rule[0ex]{0pt}{0ex} B2 & \ + Beam search & 6.11 & 15.50 & 20.15 \\ 
			\rule[0ex]{0pt}{0ex}   & \ \ + Shallow fusion &   &   &   \\ \hline \hline
			\rule[-1ex]{0pt}{3.5ex} E0 & Cascade RNN-T & 17.4 & 22.08 & 24.52 \\ \hline
			\rule[-1ex]{0pt}{3.5ex} E1 & + Convolution layer & 6.66 & 15.59 & 18.27 \\ \hline 
			\rule[-1ex]{0pt}{3.5ex} E2 & \ + Text augmentation & 6.4 & 15.24 & 17.96 \\ \hline
			\rule[-1ex]{0pt}{3.5ex} E3 & \ \ + Self shallow fusion & 5.89 & 14.93 & 17.78 \\ \hline
			\rule[-1ex]{0pt}{3.5ex} E4 & \ \ \ + Syllable correction & \textbf{5.72} & \textbf{14.85} & \textbf{17.60} \\ \hline
	\end{tabular}}
\end{table}

\begin{table}[!htb]
	\vspace{-10pt}
	\caption{Comparison of deletion, insertion and substitute errors of each model on the TA2 test set.}
	\vspace{10pt}
	\label{tab:3}
	\centering
	\scalebox{0.8}{
		\begin{tabular}{|c|l|c|}
			\hline
			\rule[-1ex]{0pt}{3.5ex} \multirow{2}{*}{\textbf{Exp ID}} & \multirow{2}{*}{\textbf{Model}} & \multicolumn{1}{c|}{\textbf{CER (\%) (D/I/S)}} \\
			\cline{3-3} \rule[-1ex]{0pt}{3.5ex} && \textbf{TA2} \\ \hline \hline
			\rule[0ex]{0pt}{0ex}   & RNN-T Character &  \\
			\rule[0ex]{0pt}{0ex} B2 & \ + Beam search &  15.50 (0.71/0.26/14.53) \\
			\rule[0ex]{0pt}{0ex}   & \ \ + Shallow fusion &   \\ \hline \hline
			\rule[-1ex]{0pt}{3.5ex} E0 & Cascade RNN-T & 22.08(0.84/0.21/21.03) \\ \hline
			\rule[-1ex]{0pt}{3.5ex} E1 & + Convolution layer & 15.59(0.76/0.17/14.66) \\ \hline 
			\rule[-1ex]{0pt}{3.5ex} E2 & \ + Text augmentation & 15.24(0.75/0.17/14.32) \\ \hline
			\rule[-1ex]{0pt}{3.5ex} E3 & \ \ + Self shallow fusion & 14.93(0.75/0.17/14.01) \\ \hline
			\rule[-1ex]{0pt}{3.5ex} E4 & \ \ \ + Syllable correction & 14.85(0.75/0.16/13.94) \\ \hline
	\end{tabular}}
	\vspace{-10pt}
\end{table}

\subsection{7500-hour Task}
We further evaluate the proposed approach on our internal large-scale 7500-hour corpus and use the same experimental setup and hyper-parameters used in AISHELL-2 Task. This time we add another more difficult test set VA that is collected at challenging acoustic conditions. As shown in Tab.~\ref{tab:4}, we first notice that with the increase of training data, there is a big CER reduction, especially on TA2 and VI test sets, as compared with the AISHELL-2 1000-hour results in Tab.~\ref{tab:2}. Comparing E9 with B5, our proposed cascade RNN-T can better enhance the language modeling ability of RNN-T than the character RNN-T with shallow fusion. Note that they both use the same 2GB text data for strengthening language modeling. In addition, on the challenging VA test set, cascade RNN-T has a large improvement over character-level RNN-T, with 14.18\% relative CER reduction. These results can prove that our proposed method can significantly improve the language modeling ability. In order to determine the advantage of using RNN-T as the S2C converter, we also try LSTM and BLSTM as the S2C converter which contains 2 LSTM or BLSTM layers with 2,048 hidden units followed by a 640-dimensional projection layer. From A0 and A1, we can observe that RNN-T based S2C converter has obvious advantages over LSTM which is also a streaming structure. The performance of RNN-T based S2C converter also surpasses the BLSTM-based one which is non-streaming and uses both past and future context.
 
\begin{table}[!htb]
	\vspace{-10pt}
	\caption{Comparison of character-level RNN-T and Cascade RNN-T on test sets in CER, training with 7500-hour corpus.}
	\vspace{10pt}
	\label{tab:4}
	\centering
	\scalebox{0.75}{
		\begin{tabular}{|c|l|c|c|c|c|}
			\hline
			\rule[-1ex]{0pt}{3.5ex} \multirow{2}{*}{\textbf{Exp ID}} & \multirow{2}{*}{\textbf{Model}} & \multicolumn{4}{c|}{\textbf{CER (\%)}} \\
			\cline{3-6} \rule[-1ex]{0pt}{3.5ex} && \textbf{TA1} & \textbf{TA2} & \textbf{VI} & \textbf{VA} \\ \hline \hline
			\rule[-1ex]{0pt}{3.5ex} B3 & RNN-T Character & 5.15 & 10.57 & 10.21 & 33.63 \\ \hline
			\rule[-1ex]{0pt}{3.5ex} B4 & + Beam search & 4.99 & 10.08 & 9.69 & 32.88 \\ \hline
			\rule[-1ex]{0pt}{3.5ex} B5 & \ + Shallow fusion & 4.85 & 9.96 & 9.5 & 32.71 \\ \hline \hline
			\rule[-1ex]{0pt}{3.5ex} E5 & Cascade RNN-T & 14.56 & 18.51 & 18.16 & 38.09 \\ \hline
			\rule[-1ex]{0pt}{3.5ex} E6 & + Convolution layer & 6.32 & 11.33 & 10.62 & 31.76 \\ \hline 
			\rule[-1ex]{0pt}{3.5ex} E7 & \ + Text augmentation & 5.53 & 10.42 & 9.59 & 29.53 \\ \hline
			\rule[-1ex]{0pt}{3.5ex} E8 & \ \ + Self shallow fusion & 4.62 & 9.33 & 8.82 & 28.21 \\ \hline
			\rule[-1ex]{0pt}{3.5ex} E9 & \ \ \ + Syllable correction & \textbf{4.57} & \textbf{9.16} & \textbf{8.65} & \textbf{28.07} \\ \hline \hline
			\rule[-1ex]{0pt}{3.5ex} A0 & RNN-T Syllable + LSTM S2C & 11.48 & 15.02 & 14.71 & 36.33\\ \hline
			\rule[-1ex]{0pt}{3.5ex} A1 & RNN-T Syllable + BLSTM S2C & 4.98 & 9.59 & 9.05 & 32.1 \\ \hline 
	\end{tabular}}
	\vspace{-10pt}
\end{table}

\subsection{Parameters, Latency and Quality}
Tab.~\ref{tab:5} summarizes several typical models in terms of the number of parameters, recognition latency and quality. Comparing E4 with B2, we can see that the proposed cascade RNN-T achieves superior performance over the character-based RNN with similar levels of model parameters and recognition latency. 

\begin{table}[!htb]
	\vspace{-10pt}
	\caption{Comparison of parameters, latency and performance for different models.}
	\vspace{10pt}
	\label{tab:5}
	\centering
	\scalebox{0.64}{
		\begin{tabular}{|c|l|c|c|c|c|c|}
			\hline
			\rule[-1ex]{0pt}{3.5ex} \multirow{2}{*}{\textbf{Exp ID}} & \multirow{2}{*}{\textbf{Model}} & \multirow{2}{*}{\textbf{Param.(M)/Latency}} & \multicolumn{4}{c|}{\textbf{CER (\%)}} \\
			\cline{4-7} \rule[-1ex]{0pt}{3.5ex} &&& \textbf{TA1} & \textbf{TA2} & \textbf{VI} & \textbf{VA} \\ \hline \hline
			\rule[0ex]{0pt}{0ex}   & RNN-T Character & & & & &   \\ 
			\rule[0ex]{0pt}{0ex} B5 & + Beam search & 93M/280ms & 4.85&9.96&9.5&32.71   \\ 
			\rule[0ex]{0pt}{0ex}   & \ + Shallow fusion & & & & &    \\ \hline \hline
			\rule[-1ex]{0pt}{3.5ex} E5 & Cascade RNN-T & 88.5M/280ms & 14.56&18.51&18.16&38.09  \\ \hline
			\rule[-1ex]{0pt}{3.5ex} E6 & + Convolution layer & 92.5M/300ms & 6.32&11.33&10.62&31.76  \\ \hline 
			\rule[-1ex]{0pt}{3.5ex} E7 & \ + Text augmentation & 92.5M/300ms & 5.53&10.42&9.59&29.53 \\ \hline
			\rule[-1ex]{0pt}{3.5ex} E8 & \ \ + Self shallow fusion & 95.5M/300ms & 4.62&9.33&8.82&28.21 \\ \hline
			\rule[-1ex]{0pt}{3.5ex} E9 & \ \ \ + Syllable correction & 95.5M/300ms & \textbf{4.57}&\textbf{9.16}&\textbf{8.65}&\textbf{28.07} \\ \hline
	\end{tabular}}\vspace{-15pt}
\end{table}

\section{Conclusions}\label{sec5}
In this paper, we propose a novel \textit{cascade RNN-T} approach to improve the language modeling ability of RNN-T. Cascade RNN-T aims to train the language model separately from the acoustic model in order to introduce a large amount of additional text to strengthen the language modeling ability. Specifically, we first use an RNN-T to transform acoustic feature into syllable sequence, and then convert the syllable sequence into character sequence by another RNN-T. By introducing several important tricks, including spanning context through convolution layer, self shallow fusion, text augmentation and syllable correction, our approach manages to surpass character-based RNN-T with a large margin on several Mandarin test sets. As our second RNN-T is an unequal mapping from pronunciation units to graphemes, we plan to try smaller units, e.g., phoneme to grapheme, particularly in English ASR in the future.


\bibliographystyle{IEEEbib}
\bibliography{refs}

\end{document}